\documentclass[graphicx]{iopart}%
\usepackage{iopams,setstack}
\usepackage{epsfig}
\providecommand{\U}[1]{\protect \rule{.1in}{.1in}}

\begin{document}

\title{Nonsequential Double Ionization with Polarization-gated Pulses}
\author{W. Quan and X. Liu}
\address{State Key Laboratory of Magnetic Resonance and Atomic and
Molecular Physics, Wuhan Institute of Physics and Mathematics,\\
Chinese Academy of Sciences, Wuhan 430071, P. R. China}
\author{C Figueira de Morisson Faria}
\address{Department of Physics and Astronomy, University College
London,\\ Gower Street, London WC1E 6BT, United Kingdom}
\begin{abstract}
We investigate laser-induced nonsequential double ionization by a
polarization-gated laser pulse, constructed employing two
counter-rotating circularly polarized few cycle pulses with a time
delay $T_{d}$. We address the problem within a classical framework,
and mimic the behavior of the quantum-mechanical electronic wave
packet by means of an ensemble of classical electron trajectories.
These trajectories are initially weighted with the quasi-static
tunneling rate, and with suitably chosen distributions for the
momentum components parallel and perpendicular to the laser-field
polarization, in the temporal region for which it is nearly linearly
polarized. We show that, if the time delay $T_{d}$ is of the order
of the pulse length, the electron-momentum distributions, as
functions of the parallel momentum components, are highly asymmetric
and dependent on the carrier-envelope (CE) phase. As this delay is
decreased, this asymmetry gradually vanishes. We explain this
behavior in terms of the available phase space, the quasi-static
tunneling rate and the recollision rate for the first electron, for
different sets of trajectories. Our results show that
polarization-gating technique may provide an efficient way to study
the NSDI dynamics in the single-cycle limit, without employing
few-cycle pulses.

\end{abstract}
\date{\today}
\maketitle

\section{Introduction}

Polarization gating, since being proposed in the mid 1990s
\cite{GatingFirst}, has turned into a powerful tool for attosecond
metrology, and dynamic imaging of matter \cite{Ottawa2008}, mainly
due to its simplicity and reliability. The key idea behind it is to
combine laser fields of different polarizations and sometimes
frequencies in such a way as to obtain controllable, near-linearly
polarized, extremely short driving laser pulses. An important
application of this technique is, for instance, the generation of
isolated attosecond pulses \cite{15,16,Chang2004}, some of them as
short as 130 attoseconds \cite{Sansone2006}. The latter pulses have
been accomplished by employing, as a driving field, a left
circularly polarized and a time-delayed right circularly polarized
sub-10fs laser pulse. The resulting driving pulse changes from
circularly polarized to almost linearly polarized, and then back to
circularly polarized, within the pulse envelope.

The importance of these drastic ellipticity changes is a consequence
of the physical mechanism behind high-order harmonic generation
(HHG). This phenomenon, which is employed in the generation of
attosecond pulses, is the result of a three-step physical process
\cite{Cork1993}. Thereby, the valence electron of an atom, subjected
to an intense laser field, is released in the continuum by tunnel
ionization. This electron is then accelerated by the field and,
depending on its energy, ionization time and on the laser-field
polarization, it may be driven back towards its parent ion and
recombine at a later instant. Upon recombination, the energy
acquired by the electron from the field is released as a
high-frequency photon. Gating is important, as it may be employed to
restrict the electron ionization and recombination times to very
narrow intervals. If the field is circularly polarized, the
returning electron will miss the atomic core, whereas, if the field
is linearly polarized, the electron will interact with it.

Laser-induced nonsequential double ionization (NSDI) is also caused
by a similar mechanism. The main difference lies on the fact that
the returning electron, instead of recombining with a bound state of
its parent ion, will rescatter inelastically with it, giving part of
its kinetic energy to release a second electron. This nonsequential
physical mechanism has been revealed by very peculiar features in
the electron momentum distributions, namely symmetric peaks at the
nonvanishing momenta $p_{1\parallel}=p_{2\parallel}=$
$\pm2\sqrt{U_{p}}$, where $U_{p}$ denotes the ponderomotive energy
and $p_{n\parallel}(n=1,2)$ the electron momentum components
parallel to the laser field polarization \cite{NSDIexp}. Therefore,
we expect that NSDI may also be strongly influenced by the time
dependence of the external laser field, for example, by the dramatic
change of its amplitude and polarization. This is supported by our
previous work on NSDI with few-cycle, linearly polarized driving
pulses \cite{fewcycleexp,LF2004,FLSL2004}. Therein, we have shown,
both theoretically \cite{LF2004,FLSL2004} and experimentally
\cite{fewcycleexp}, that the NSDI dynamics are strongly dependent on
the carrier-envelope (CE) phase. Indeed, we observed strongly
asymmetric distributions, which would shift from the positive to the
negative region, or vice versa, of the plane $(p_{1\parallel
},p_{2\parallel})$ spanned by the electron momentum components
$p_{n\parallel }(n=1,2)$ parallel to the laser-field polarization,
as the CE phase was changed. Furthermore, it has been shown, for
NSDI with a single elliptically polarized driving field, that the
electron-momentum distributions are highly dependent on the
driving-field polarization \cite{30}.

Hence, one can raise the question of whether polarization gating may
be used to control NSDI dynamics. This issue is of interest for two
main reasons. Firstly, since one may build an extremely short,
almost linearly polarized driving field, it may be possible to steer
the electron motion with a much higher precision than with, for
instance, linearly polarized few-cycle pulses. Secondly, sometimes
the changes observed in the NSDI electron momentum distributions,
with regard to the driving-field shape, are far more dramatic than
in high-order harmonic or above-threshold ionization spectra. This
is due to the fact that different sets of electron orbits may be
mapped into different momentum regions. Apart from that, if the
second electron is dislodged by electron-impact ionization, there
will be a minimum and a maximum parallel momentum for which this
process is classically allowed. Depending on the field parameters,
we can employ this particular feature to make whole momentum regions
appear or collapse. The above-stated effects have been observed for
NSDI with linearly polarized few cycle pulses, and led to dramatic
changes in the electron momentum distributions
\cite{fewcycleexp,LF2004,FLSL2004}.

In this work, we investigate differential electron momentum
distributions in NSDI with polarization gating. For that purpose, we
extend the classical model employed in our previous publications
\cite{LF2004,FLSL2004} to the elliptical-polarization case. In this
model, the quantum-mechanical transition probability corresponding
to the scenario in which the second electron is released by
electron-impact ionization, within the strong-field approximation,
is mimicked by a classical ensemble of electrons released in the
continuum with a quasi-static tunneling rate
\cite{LF2004,FLSL2004,FSLB2004}. The main difference is that, for
elliptically polarized fields, one must take into account that the
lateral residual laser electric field influences the electron orbits
in the continuum, and also at the instant of ionization. As the
external driving field, we consider the same pulse configuration as
in \cite{Chang2004}, i.e., two counter-rotating, time-delayed
circularly polarized few-cycle pulses. We investigate the influence
of both the CE phase and of the delay between the two pulses on the
electron momentum distributions. In a more general context, it is
worth mentioning that classical models have proven to be very
powerful in the context of nonsequential double \cite{classicalNSDI}
or multiple \cite{classicalNSMI} ionization.

This paper is organized as follows. In Sec. \ref{model}, we provide
a brief discussion of our model, placing a particular emphasis on
how it differs from its counterpart for linearly polarized fields.
Subsequently (Sec. \ref{results}), we present the differential
electron momentum distributions, and analyze their main features in
terms of electron trajectories. Finally, our conclusions and a
summary of this work are given in Sec. \ref{conclusions}.

\section{Model}

\label{model}

We consider an electron ensemble subject to a pulse with a
time-dependent ellipticity, which is generated by the superposition
of a left-circularly polarized pulse, and right-circularly polarized
pulse. The two pulses are taken to be identical except by their
polarization, and there is a time delay $T_{d}$ between them. Below,
we provide more details about the pulse shape, and our classical
ensemble model, which is employed to imitate the behavior of a
quantum-mechanical wave packet. Atomic units are being used
throughout.

\subsection{Polarization-gated driving field}

\label{pulseshape}

Explicitly, the electric fields $\overrightarrow{E}_{l}(t)$ and
$\overrightarrow{E}_{r}(t)$ of the left- and right circularly polarized pulses
read
\begin{eqnarray}
\overrightarrow{E}_{l}(t)  &  =E_{0}e^{-2\ln(2)((t-T_{d}/2)/\tau_{p})^{2}%
}\nonumber\\
&  \lbrack\hat{x}\cos(\omega(t-T_{d}/2)+\phi)+\hat{y}\sin(\omega
(t-T_{d}/2)+\phi)] \label{eq1}%
\end{eqnarray}%
\begin{eqnarray}
\overrightarrow{E}_{r}(t)  &  =E_{0}e^{-2\ln(2)((t+T_{d}/2)/\tau_{p})^{2}%
}\nonumber\\
&  [\hat{x}\cos(\omega(t+T_{d}/2)+\phi)-\hat{y}\sin(\omega(t+T_{d}/2)+\phi)],
\label{eq2}%
\end{eqnarray}
respectively. In the above-stated equations, $E_{0}$ is the
peak-field amplitude, $\omega$ is the carrier frequency, $\tau_{p}$
is the pulse duration, $T_{d}$ is the time delay between the two
circularly polarized pulses and $\phi$ is CE phase. The unit vectors
in the $x$ and $y$ directions are denoted by $\hat{x}$ and
$\hat{y}$.

The electric field components of the combined laser pulse in the $x$
and $y$ direction are given by
\begin{eqnarray}
\overrightarrow{E}_{x}(t)  &  =E_{0}e^{-2\ln(2)((t-T_{d}/2)/\tau_{p})^{2}%
}[\cos(\omega (t-T_{d}/2)+\phi)]\nonumber\\
&
+E_{0}e^{-2\ln(2)((t+T_{d}/2)/\tau_{p})^{2}}[\cos(\omega(t+T_{d}/2)+\phi)]
\label{eq3}%
\end{eqnarray}
and%
\begin{eqnarray}
\overrightarrow{E}_{y}(t)  &  =E_{0}e^{-2\ln(2)((t-T_{d}/2)/\tau_{p})^{2}%
}[\sin(\omega (t-T_{d}/2)+\phi)]\nonumber\\
&  -E_{0}e^{-2\ln(2)((t+T_{d}/2)/\tau_{p})^{2}}[\sin(\omega(t+T_{d}/2)+\phi)],
\label{eq4}%
\end{eqnarray}
respectively. The time-dependent ellipticity of this pulse is
\begin{equation}
\xi(t)=\frac{|1-\exp[-4\ln(2)tT_{d}/\tau_{p}^{2}]|}{1+\exp[-4\ln(2)tT_{d}%
/\tau_{p}^{2}]}.
\end{equation}
In the vicinity of $t=0,$ $\xi(t)$ increases from 0 to 0.2 and is
approximately linear. Outside this interval, this approximation does not hold.
The temporal region for which the field polarization is approximately linear
is known as the ``polarization gate" \cite{GatingFirst}.

In Ref. \cite{Chang2004}, the polarization gate for the specific
laser-field configuration discussed above has been estimated to be
around $0.3\tau_{p}$. Therein, it has also been shown that it is
inversely proportional to the time delay $T_{d}$ between the two
pulses. Hence, to reduce the polarization gate, we can either use
shorter circularly polarized pulses or increase $T_{d}$. From the
experimental perspective, either of them can be controlled at will.
In fact, the length of the laser pulse will be conditioned by
state-of-the-art ultrafast laser technique \cite{17}. However, the
delay time between the two circularly polarized laser pulses cannot
be too long. If we use a too long delay between the two circularly
polarized laser pulses, to obtain a sufficiently intense combined
pulse it may be necessary to strengthen our laser pulses so much
that the ground-state sample may be depleted and result in a poor
signal-to-noise ratio.

\subsection{Electron momentum distributions}

We will now discuss how we mimic the quantum-mechanical electron
momentum distributions in a classical framework. Similar models have
been employed in \cite{LF2004,FLSL2004,FSLB2004} for linearly
polarized fields. We consider a set of classical trajectories,
starting at different tunneling times $t_{0}$ throughout the pulse.
We limit such times to the time interval for which the field is
almost linearly polarized. This is justified, as, only in this time
range, a significant contribution to NSDI will occur. Furthermore,
in the specific model discussed in this paper, we make three main
assumptions. First, each trajectory is weighted with the tunneling
probability per unit time given by the well-known quasi-static
formula \cite{6}
\begin{equation}
W_{t}(t_{0})\thicksim\frac{1}{|E_{\mathrm{sum}}(t_{0})|}\exp\left[
\frac{-2(2|E_{IP1}|)^{3/2}}{3|E_{\mathrm{sum}}(t_{0})|}\right]  , \label{eq7}%
\end{equation}
where $E_{IP1}$ is the first ionization threshold of the atom in question, and
$\vec{E}_{\mathrm{sum}}(t_{0})$ $=\overrightarrow{E}_{x}(t_{0}%
)+\overrightarrow{E}_{y}(t_{0})$ is the electric field of the
combined, polarization-gated laser pulse. Second, to simulate the
initial wave-packet spreading, an initial lateral velocity $v_{l}$
is further introduced. Each trajectory is then weighted with the
tunneling probability times the quantum-mechanical transverse
velocity distribution weight \cite{8}, $W_{l}$, which can be
calculated by
\begin{equation}
W_{l}(v_{l})=\frac{1}{\pi(\delta{v_{l}})^{2}}\exp[-(\frac{v_{l}}{\delta{v_{l}%
}})^{2}], \label{eq8}%
\end{equation}
where the lateral velocity width is given by
\begin{equation}
{\delta}v_{l}=(E_{\mathrm{sum}}/\sqrt{2E_{IP1}})^{1/2}. \label{eq9}%
\end{equation}
The range for the transverse velocity distribution chosen here is
$2{\delta }v_{l}$. Finally, a quantum-mechanical wave packet also
spreads in time, so that the contributions of the longer orbits are
weakened. We incorporate this final ingredient in a similar way as
in Ref. \cite{30}, by introducing an extra weight
$(t_{1}-t_{0})^{-3}$ to each electron trajectory. Thereby, $t_{1}$
denotes the time the electron returned to the core. Therefore, the
overall weight of each trajectory in the ensemble reads
\begin{equation}
W_{k}(t_{0},v_{l})=W_{t}(t_{0}){\times}W_{l}(v_{l}){\times}(t-t_{0})^{-3}.
\label{eq10}%
\end{equation}
After tunneling, the equations of motion for the electron in both directions
are
\begin{equation}
\ddot{x}=\overrightarrow{E}_{x}(t) \label{eq5}%
\end{equation}%
\begin{equation}
\ddot{y}=\overrightarrow{E}_{y}(t). \label{eq6}%
\end{equation}
In order to obtain the electron orbits, Eq. (\ref{eq5}) and Eq. (\ref{eq6})
are integrated, and the Coulomb potential is ignored. This is a reasonable
assumption for a strong driving field. Electrons are assumed to be `born' at
time $t_{0}$ at the origin $x=0,y=0$ with initial lateral velocity. Only those
electrons coming back to the core with energy larger than $E_{IP2}$, the
second ionization threshold of the atom, will contribute to the NSDI yield.
This corresponds to the physical situation in which the first electron
dislodges the second by electron-impact ionization. After a trajectory is
launched, by checking the position of the electron, we determine whether it
has returned to the core, and, if so, its return time. The electron velocities
in both directions, $\dot{x}$ and $\dot{y}$, are then evaluated. The energy of
the electron, upon return, is
\begin{equation}
E_{ret}=\frac{1}{2}(\dot{x}^{2}+\dot{y}^{2}). \label{eq11}%
\end{equation}
If the condition $E_{ret}>E_{IP2}$ is satisfied \cite{FSLB2004}, the first
electron gives part of its kinetic energy $E_{ret}$ upon return to a second
electron, so that it is able to overcome the second ionization threshold
$E_{IP2}$. The electron pair then obeys
\begin{eqnarray}
\frac{1}{2}\sum_{j=1}^{2}[p_{jx}+A_{x}(t_{1})]^{2}  &  =E_{\mathrm{ret}%
}-E_{IP2}-\frac{1}{2}\sum_{j=1}^{2}p_{jz}^{2}\nonumber\\
&  -\frac{1}{2}\sum_{j=1}^{2}[p_{jy}+A_{y}(t_{1})]^{2}. \label{eq13}%
\end{eqnarray}
In Eq. (\ref{eq13}), $A_{x}(t)$ and $A_{y}(t)$ denote the vector
potential-components
\begin{equation}
A_{x}(t)=-{\int}E_{x}(t)dt \label{eq14}%
\end{equation}%
\begin{equation}
A_{y}(t)=-{\int}E_{y}(t)dt \label{eq16}%
\end{equation}
of the laser pulse in the $x$ and $y$ directions, and $p_{jx}$,
$p_{jy}$ and $p_{jz}$ are the final momentum components recorded by
the detector in the $x$, $y$ and $z$ direction after tunneling.

One should note that the energy-conservation condition (Eq.
(\ref{eq13})) gives the equation of a hypersphere in the
six-dimensional space spanned by the momentum components of the two
electrons. This hypersphere exhibits the radius
$[2(E_{\mathrm{ret}}(t_{1})-E_{IP2})]^{1/2}$ and is centered at $(p_{1x}%
,p_{1y},p_{1z};p_{2x},p_{2y},p_{2z})=$ $(-A_{x}(t_{1}),-A_{y}(t_{1}%
),0;-A_{x}(t_{1}),-A_{y}(t_{1}),0).$ Therefore, the larger the
electron kinetic energy is, the larger is the region for which
electron-impact ionization is classically allowed. Furthermore, the
above-stated equation shows that the distributions should be
centered at nonvanishing electron momenta. Specifically for
monochromatic, linearly polarized fields, the vector potential upon
return can be approximated by $2\sqrt{U_{p}}.$ This corresponds to
the situation in which the electron leaves at a field maximum and
returns at a crossing. Since this is the most probable momentum for
the electron upon return, we expect the distributions to be centered
at this quantity. We have observed, however, for the parameters
employed in this work, that this estimate roughly holds.

The electron momentum distributions then read
\begin{eqnarray}
R  &  {\thicksim}{\int}{\int}dt_{0}dv_{l}W_{k}(t_{0},v_{l})\nonumber\\
&  {\delta}(E_{\mathrm{ret}}(t_{1})-E_{IP2}-\sum_{j=1}^{2}\frac{[p_{jx}%
+A_{x}(t_{1})]^{2}}{2}-\nonumber\\
&  \sum_{j=1}^{2}\frac{[p_{jy}+A_{y}(t_{1})]^{2}}{2}-\sum_{j=1}^{2}%
\frac{p_{jz}^{2}}{2}). \label{eq15}%
\end{eqnarray}
The argument of the $\delta$ function gives the energy-conservation
restriction. Since the situation addressed in this paper occurs only
in the $xy$ plane, the electron motion in the $z$ direction is
ignored.

In Eq. (\ref{eq15}), there is also an additional assumption, namely
that the second electron is released by a contact-type interaction
placed at the position of the ion. In this case, the electron
momentum distributions are mainly determined by the momentum-space
integration. This means that they are isotropic in momentum space,
have the radius of the above-mentioned hypersphere, and are centered
at the most probable momentum upon return. This guarantees that we
investigate only the effects of the gating on the NSDI momentum
distributions. Other types of electron-electron interaction would
lead to a term $|V_{\mathbf{p}_{n}\mathbf{k}}|^{2}$ dependent on the
final momenta $\mathbf{p}_{n}(n=1,2)$, and on the intermediate
momentum $\mathbf{k}$ in the above distributions (for details see
Ref. \cite{FSLB2004}). Such extra momentum dependence might distort
or mask the effects of the polarization gating.

\section{Results}
\label{results}

\begin{figure}[ptb]
\begin{center}
\includegraphics[width=4.5in]{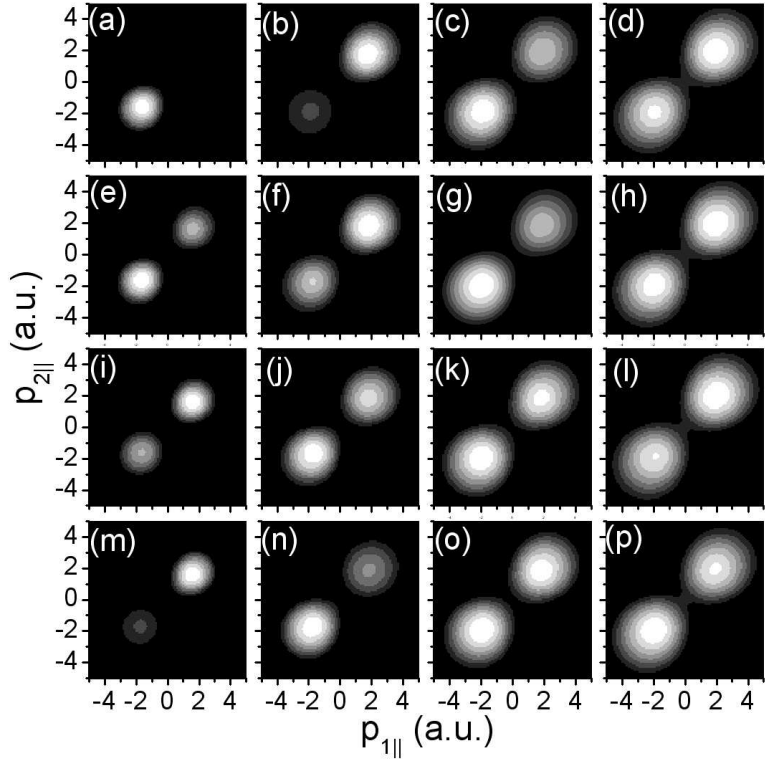}
\end{center}
\caption{NSDI electron momentum distributions computed for Neon in a
polarization-gated pulse described by Eqs. (3)-(4). The peak
intensity of the two circularly polarized pulses is
$2\times10^{14}\mathrm{W/cm}^{2},$ their length is four cycles
($n=4$) and their frequency is \ $\omega=0.057$ a.u. The
distributions are plotted as functions of the momentum components
$p_{n\parallel}(n=1,2)$ parallel to the nearly linearly polarized
part of the combined pulse. The CE phase is varied from the top to
the bottom of the figure, and the delay $\omega T_{d}$ between the
two pulses from its left to its right. In the first, second, third
and fourth rows from the top [panels (a) to (d), panels (e) to (h),
panels (i) to (l), and (m) to (p), respectively], the CE phases
$\phi$ are $0.5\pi,$ $0.8\pi ,1.0\pi,$ and $1.2\pi$ respectively. In
the first [panels (a), (e), (i) and (m)], second [panels (b), (f),
(j) and (n)], third [panels (b), (f), (k) and (o)], and fourth
[panels (c), (g), (l) and (p)] columns from the left, we depict
distributions for the delay phases $\omega T_{d}=8\pi,$ $6\pi,4\pi,$
and $2\pi$.}%
\label{fig1}%
\end{figure}In this section, we display the NSDI electron-momentum
distributions computed with Eq. (\ref{eq15}) and the
polarization-gated pulse described in Sec. \ref{pulseshape}. This
pulse has been chosen as the superposition of two time-delayed,
circularly polarized pulses of wavelength $\lambda=800$ nm. The full
width at half maximum (FWHM) of the circularly polarized light is
chosen as four cycles, i.e., $\tau \sim$ 10 fs. The distributions
were calculated for neon. For this specific species, the dominant
physical mechanism is electron-impact ionization. For other species,
such as argon, the excitation of the parent ion by the returning
electron, with subsequent double ionization, plays an important role
\cite{NSDIexp,jesus2004}.

Such distributions are depicted in Figure \ref{fig1}, as functions
of the electron momentum components parallel to the laser-field
polarization at the very center of the gate. For the pulse shape
considered in this paper, this corresponds to the $x$ direction.
Unless otherwise stated (see, e.g., Figs. 5 and 6), we will denote
such momentum components as $p_{n\parallel}(n=1,2)$. One should
note, however, that, in the calculations, the electron is propagated
in both $x$ and $y$ directions.

If the delay phase between both pulses is equal to the pulse length,
i.e., $\omega T_{d}=8\pi,$ the distributions are, in general, highly
asymmetric and concentrated either in the positive or the negative
parallel momentum regions. For the specific parameters in this work,
as the CE phase $\phi$ is changed from $\phi=0.5\pi$ to
$\phi=1.2\pi,$ the distributions shift from the third to the first
quadrant of the plane $(p_{1\parallel},p_{2\parallel})$ spanned by
the parallel-momentum components. This behavior is shown in the far
left panels of Fig. \ref{fig1}, i.e., in Figs. \ref{fig1}(a),
\ref{fig1}(e), \ref{fig1}(i), and \ref{fig1}(m), and resembles to a
great extent what happens for linearly polarized few-cycle pulses.

In this latter case, we have shown that the asymmetry and the shifts
in the distributions were due to the changes in the dominant set of
trajectories along which the first electron would return. This set
of trajectories strongly depends on the CE phase, so that the
asymmetry can be used to determine this parameter
\cite{LF2004,FLSL2004}. The behavior of the electron-momentum
distributions in the present case suggests a similar physical
interpretation. From the positions of maxima, at approximately
$(2\sqrt{U_{p}},2\sqrt{U_{p}})$ and
$(-2\sqrt{U_{p}},-2\sqrt{U_{p}})$, we verify that the estimates for
the peaks of the distributions, for monochromatic, linearly
polarized fields, roughly holds in this case.

As the delay phase is reduced to $\omega T_{d}=6\pi$, the
distributions become wider and the electron yield becomes much
higher. This is shown in the second column of Fig.~\ref{fig1}
[Figs.~1(b), 1(f), 1(j) and 1(n)]. The increase in width suggests
that the radius of the hypersphere (\ref{eq13}), which delimits the
classically allowed region, increased. Physically, this would
correspond to a larger kinetic energy $E_{\mathrm{ret}}(t_{1})$ of
the first electron upon return. The growth in electron yield hints
at an increase in the tunneling probability for the first electron.
The distributions, however, exhibit a similar qualitative behavior
to that observed for $\omega T_{d}=8\pi$, in the sense that they are
asymmetric and depend on the CE phase.

A closer inspection, however, reveals a dramatic change, in the
sense that the electron distributions are concentrated in an
opposite momentum region, as compared to their counterparts at the
delay phase of $8\pi$.  For example, if the CE phase is $0.5\pi$,
such distributions are almost entirely localized in the first
quadrant of the parallel momentum plane for  $\omega T_{d}=6\pi$
[Fig. 1(b)], while, for $\omega T_{d}=8\pi$, they occupy the
negative momentum region [Fig. 1(a)]. This sharp contrast persists
for the other values of CE phases. The reason behind this shift will
be addressed later.

If the delay between both pulses is decreased further, the overall
asymmetry in the electron momentum distributions starts to fade.
This may be seen in the third column of Fig.~1 [Figs.~1(c), 1(g),
1(k) and 1(o)], for which $\omega T_{d}=4\pi$. Such distributions
are only slightly asymmetric and exhibit bright spots both in the
first and third quadrants of the $(p_{1\parallel},p_{2\parallel})$
plane. This trend persists for an even shorter delay of $\omega
T_{d}=2\pi$, as shown in the far right panels of Fig.~1 [Figs.~1(d),
1(f), 1(l), and 1(p)]. In this latter case, the distributions are
nearly symmetric and the information about the CE phase is almost
lost.

This result is rather counterintuitive, as it resembles more the distributions
obtained for a linearly monochromatic driving field than what is expected for
a combination of two few-cycle pulses. In the former case, the distributions
are totally symmetric upon $(p_{1\parallel},p_{2\parallel})\rightarrow
(-p_{1\parallel},-p_{2\parallel})$. In our previous work \cite{LF2004}, we
observe that this symmetry holds in practice if the pulse is longer than 10
cycles. However, the longest quasi-linearly polarized pulse considered here
has no more than 4 cycles, as the gate width must be less than the laser-pulse
duration. Therefore, intuitively, one would anticipate asymmetric
distributions. In the following, we will give an explanation for such a
surprising result based on lateral electron dynamics.

\begin{figure}[ptb]
\begin{center}
\includegraphics[width=4.0in]{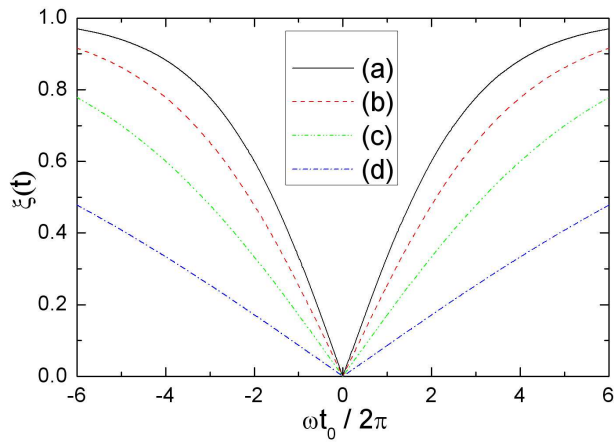}
\end{center}
\caption{The ellipticity of the combined field
$\overrightarrow{E}_{\mathrm{sum}}$, close to the center of the
polarization gate, as a function of the electric field initial phase
$\omega t_0$. The CE phase of the laser field is chosen as $0.5\pi$,
while the delay phases are $8\pi$, $6\pi$, $4\pi$ and $2\pi$ (tags
(a), (b), (c) and (d), respectively). The remaining parameters are
the same as in the previous figure.} \label{fig2}
\end{figure}

For that purpose, we will have a closer look at how the ellipticity
of the combined laser field varies, with respect to the delay phase
$\omega T_{d}$. These results are displayed in Fig.~2, and show that
the delay phase $\omega T_{d}$ has a strong influence on the
ellipticity of the very-center portion of the combined laser field.
A longer delay phase corresponds to a sharper slope. When $\omega
T_{d}=8\pi$, the ellipticity drops very steeply to zero around
$\omega t_{0}=0$ and changes back soon after that. Hence, for longer
time delays $T_{d}$, the start times $t_{0}$ of the first electron
are restricted to a very narrow interval and the gate behaves
better. This sheds some light on why the results in the first column
in Fig. \ref{fig1}, computed for $8\pi$ delay phase, are similar to
those obtained with linearly polarized driving fields. In contrast,
we have verified that variations in the CE phase do not change the
ellipticity of the combined laser field.

\begin{figure}[ptb]
\begin{center}
\includegraphics[width=5in]{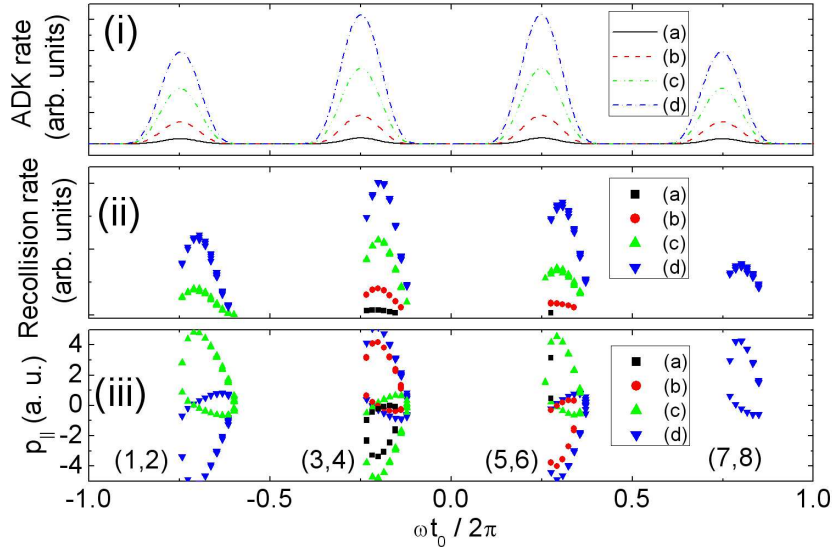}
\end{center}
\caption{Ionization and recollision rates, together with the final
electron momentum along $p_{1||}=p_{2||}=p_{||}$, as functions of
the initial phase $\omega t_{0}$, for a pulse described by Eqs.
(\protect{\ref{eq3}}) and (\protect{\ref{eq4}}), with CE phase
$\phi=0.5\pi$. Panel (i) exhibits the calculated ADK ionization
rate. Panel (ii) displays the final recollision rate, calculated
with the ADK rate and lateral velocity distribution rate. Panel
(iii) gives the final momentum $p_{||}$. The tabs (a),(b),(c),(d)
correspond to the delays $\omega T_{d}=8\pi$, $6\pi$, $4\pi$ and
$2\pi$, respectively. The numbers $(j,\nu)$ in panel (iii) indicate
a pair of electron orbits. The remaining parameters are the same as
in Fig.~\ref{fig1}. } \label{fig3}
\end{figure}

In Fig.~\ref{fig3}, we analyze the behavior of the ionization rate
for the first electron and the phase-space contributions with
respect to delay phase $\omega T_{d}$. We consider parallel momenta
along the diagonal, i.e.,
$p_{1\parallel}=p_{2\parallel}=p_{\parallel}$, and impose that the
perpendicular momentum components vanish. The classically allowed
momentum region is most extensive for vanishing transverse momenta.
Hence, we expect that the figure will provide a rough upper bound
for it. We also fix the CE phase at $0.5\pi$. A similar analysis has
been performed in Ref. \cite{LF2004} for linearly polarized
few-cycle pulses.

In the two upper panels [Fig.~\ref{fig3}(i) and Fig.~\ref{fig3}(ii),
respectively], we display the calculated ADK ionization rate in the
central portion of the combined pulse, and the final recollision
rate for the first electron. The tabs, (a),(b),(c),(d) in the figure
follow those in Fig. \ref{fig1}, i.e., from (a) to (d), the delay
phase is varied from $8\pi$ to $2\pi$. The recollision rate is
weighted according to Eq. (\ref{eq8}) and Eq.~(\ref{eq10}) for the
initial lateral velocity distribution. The recollision rate is
weighted according to Eq. (\ref{eq10}) and decided by three factors,
i.e., the tunneling ionization rate, the wave packet spread time and
the quantum-mechanical transverse velocity distribution weight.
Since the lateral electric field of the combined laser pulse will
shift the electron wave packet laterally, trajectories with zero
initial lateral velocity, which may return to the core in a linearly
polarized laser, will miss the core. Only those with a certain
initial lateral velocity, which may compensate the displacement
induced by the lateral electric field, can return to the core and
contribute to the NSDI yield. The effect of the transverse velocity
has been demonstrated in HHG by a recent experiment where an
elliptical laser pulse was employed \cite{26}.

>From Eq.~(\ref{eq9}), one can see that, as compared to the
trajectories with vanishing initial lateral velocity, those with a
non-vanishing initial lateral velocity have a lower transverse
velocity distribution weight, which in turn reduces the final
recollision rate. Therefore, the recollison rate will be corrected
according to the transverse electron dynamics. Comparing
Figs.~\ref{fig3}(i) and (ii), one also finds that the highest
recollision rate appears slightly later than the peak of the ADK
ionization rate. This is a consequence of the fact that the
trajectories starting near the peak laser electric field will
propagate with longer times and be affected more by the wave-packet
spreading effect. This effect results in the drop of the recollision
rate according to a cubic power law with the evolution time
$t_1-t_0$.

The parallel momenta $p_{\parallel}$ in Fig.~\ref{fig3}(iii) provide
an approximate estimate for the region in momentum space for which
electron-impact ionization is classically allowed. The region
delimited by such momenta gives a very good idea of the role of
phase-space effects: the larger this region is, the more important
they are. A small region, on the other hand, means a small radius
for the hypersphere in Eq. (\ref{eq15}), which, physically,
indicates that the second electron may only be dislodged in a small
region in momentum space. In Fig.~\ref{fig3}(iii), one may identify
a few sets of orbits, which lead to parallel momenta either in the
positive or negative region. Starting from the left, these orbits
are denoted by $(1,2)$, $(3,4)$, $(5,6)$ and $(7,8)$.

The momentum region to which they contribute depend on the time
delay $\omega T_d$. If this delay is an even multiple of the field
cycle $2\pi$, i.e., for $\omega T_d=4\pi$ and $\omega T_d=8\pi$,
there exist at most two sets of orbits, $(1,2)$ and $(5,6)$, which
yield contributions in the first quadrant of
$(p_{1\parallel},p_{2\parallel})$.  These orbits start within
$-2\pi<\omega t_{0}<-\pi$ and $0<\omega t_{0}<\pi$, respectively.
The remaining sets of orbits, starting at $0<\omega t_{0}<\pi/2$ and
$3\pi/2<\omega t_{0}<2\pi$, lead to negative parallel momenta. If
$\omega T_d$ is an odd multiple of the field cycle, i.e., for
$\omega T_d=6\pi$ and $\omega T_d=2\pi$, the above-mentioned
situation is reversed.

For linearly polarized driving fields, the ADK ionization rate and
phase-space effects suffice, in order to determine whether a set of
orbits contributes significantly to the electron-momentum
distributions. A large rate indicates that the first electron will
tunnel with a significant probability per unit time. Furthermore, if
this electron returns with sufficient energy to release the second
electron over a significant region of momentum space, one expects
that the contributions from a specific set of orbits will be
prominent \cite{LF2004,FLSL2004}. For a polarization-gated pulse,
however, we have to take into account the lateral electron dynamics.
This information is embedded in the recollision rate depicted in
Fig. \ref{fig3}(ii).

We will now analyze the interplay of the above-mentioned issues in
the electron momentum distributions. For a delay $\omega
T_{d}=8\pi$, there is mainly a single set $(3,4)$ of trajectories
for which electron-impact ionization is classically allowed. This
set corresponds to $\omega t_{0}\sim-\frac{\pi}{2}$ and leads to
contributions in the negative momentum region. The contribution from
another set $(5,6)$ of orbits may be ignored due to its small
classical allowed region and low recollision rate.  For this reason,
the distributions are highly asymmetric and concentrated in the
third quadrant of the parallel-momentum plane $(p_{1\parallel},
p_{2\parallel})$, in agreement with Fig.~\ref{fig1}(a). This very
restricted range in the classically allowed region is possibly due
to the fact that the ellipticity of the driving field changes very
fast. Hence, the first electron only returns to its parent ion
within a very narrow temporal region. The tunneling rate is also
weak in this case, as it is taken only in the central part of the
combined pulse and there is very little overlap between each
circularly polarized pulse for this specific delay.

As the delay phase decreases to $\omega T_{d}=6\pi$, the situation
becomes different. In this case, the orbits $(3,4)$ lead to
$p_{\parallel}>0$. Hence, the distributions are concentrated in the
positive parallel momentum region. Apart from that, there is also a
further set $(5,6)$ of orbits, whose start times lie near $\omega
t_0 \sim \frac{\pi}{2}$. The contributions of this latter pair are
weaker, and localized in the third quadrant of the plane
$(p_{1\parallel},p_{2\parallel})$. Finally, as an overall feature,
there is an increase in the ADK rate and also in the recollision
rate for the first electron. This is due to an increase in the
overlap between the two pulses $\overrightarrow{E}_{r}$ and
$\overrightarrow{E}_{l}$, and leads to brighter distributions. All
the above-stated features can be observed in Fig.~\ref{fig1}(b).

A further reduction in the delay phase to $\omega T_{d}=4\pi$ leads
to an additional set $(7,8)$ of orbits for which the first electron
may return and release the second electron. This set starts near
$\omega t_0 \sim -3 \pi/2$, and leads to contributions in the first
quadrant of the plane $(p_{1\parallel},p_{2\parallel})$. This will
add up to the contributions from the orbits $(5,6)$. Hence, overall,
there will be two sets of orbits yielding momenta in such a region.

The ionization and recollision rates, however, are larger for the
orbits $(3,4)$ starting at $\omega t_{0} \sim -\pi/2$. Therefore,
the distributions are slightly brighter in the negative momentum
region. Finally, for $\omega T_{d}=2\pi$, there are four sets of
orbits contributing to the electron-momentum distributions, and the
distributions are approximately symmetric.

\begin{figure}[ptb]
\begin{center}
\includegraphics[width=5in]{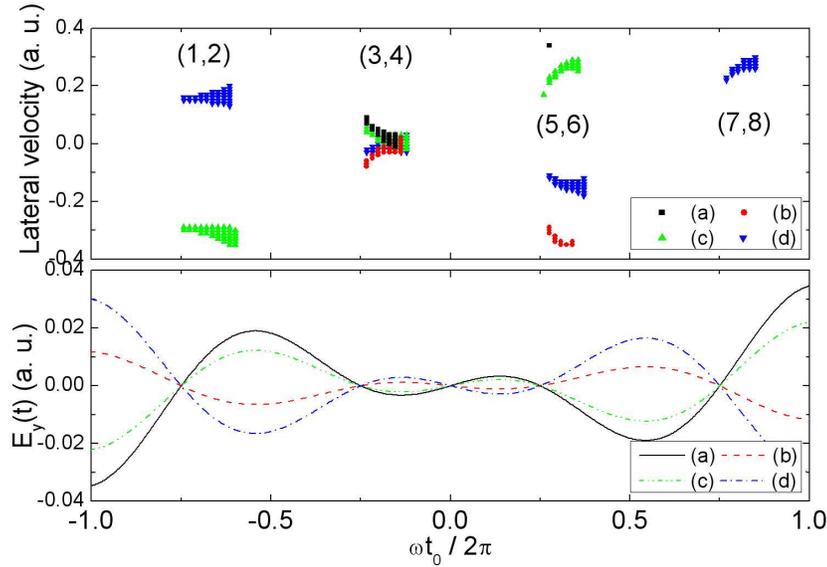}
\end{center}
\caption{The upper and lower panel show the initial lateral
velocities of the rescattering orbits and the lateral laser electric
field, respectively, as functions of the laser electric field phase
$\omega t_{0}$. The tabs, (a),(b),(c),(d), follow those in Fig.
\ref{fig3}. The numbers $(i,j)$ in the upper panel denote the orbits composing a pair.}%
\label{fig4}%
\end{figure}

In Fig.~\ref{fig4}, we display the initial lateral velocity for the
rescattering orbits (upper panel) and the lateral laser electric field (lower
panel), as functions of the initial electric field phase $\omega t_{0}$.
Classically, for a linearly polarized driving electric field, an electron with
a large lateral initial velocity will miss the core. For an elliptically
polarized field, however, the lateral electric field will also slightly change
the electron's orbit in the transverse direction.

If both effects are combined, a gate can be formed to choose the
orbit with a certain initial lateral velocity to come back with
highest probability \cite{26}. A larger lateral velocity corresponds
to a stronger lateral laser electric field and, according to
Eq.~(\ref{eq8}), to a lower ionization rate. In Fig.~\ref{fig4}, the
orbits starting near $\omega t_{0}\sim-\frac{\pi}{2}$ meet the
smallest lateral laser electric field because their tunneling phases
are those nearest to the center of the gate. Hence, these orbits
exhibit the largest recollision probability. This explains why we
observe a favored NSDI rate in this region in Fig. \ref{fig3}.

It is also noteworthy that this specific set of orbits exhibits lateral
velocities close to zero, whereas the lateral velocities of the remaining sets
vary considerably with the time delay. In general, as the delay phase $\omega
T_{d}$ decreases, the lateral velocities diminish as well. This causes an
overall increase in the tunneling and recollision rates, in agreement with
Fig.~\ref{fig3}.

Apart from that, specifically for the time delay $\omega
T_{d}=2\pi$, there is an overall decrease in the lateral velocities
for the orbits starting near $\omega t_{0} \sim-\pi$ and $\omega
t_{0} \sim\pi$. Both sets of orbits lead to positive final momenta
$p_{n\parallel}(n=1,2)$. Thus, we will expect an increase in
brightness in the first quadrant of the $(p_{1\parallel
},p_{2\parallel})$ plane. As a direct consequence, the distributions
are nearly symmetric. The set of orbits $(7,8)$, starting near
$\omega t_{0}=1.5\pi$, exhibit very large transverse velocities and
therefore does not contribute significantly to the yield.

\begin{figure}[ptb]
\begin{center}
\includegraphics[width=6in]{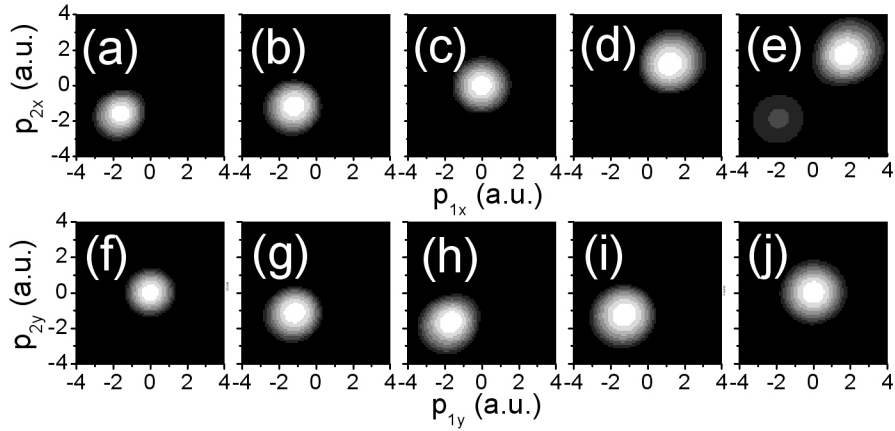}
\end{center}
\caption{NSDI electron momentum distributions computed for neon in a
polarization-gated pulse. The upper and lower panels give the
distributions as functions of the momentum components $(p_{1x},
p_{2x})$ and $(p_{1y},p_{2y})$, respectively. The CE phase is fixed
to be $0.5\pi$ and the delay phases, from left to right, are $8\pi$
[panels (a) and (f)], $7.5\pi$ [panels (b) and (g)], $7\pi$ [panels
(c) and (h)], $6.5\pi$ [panels (d) and (i)] and $6\pi$ [panels (e)
and (j)].
The remaining parameters are the same as in Fig. \ref{fig1}.}%
\label{pd_odddelay}%
\end{figure}

\begin{figure}[ptb]
\begin{center}
\includegraphics[width=5in]{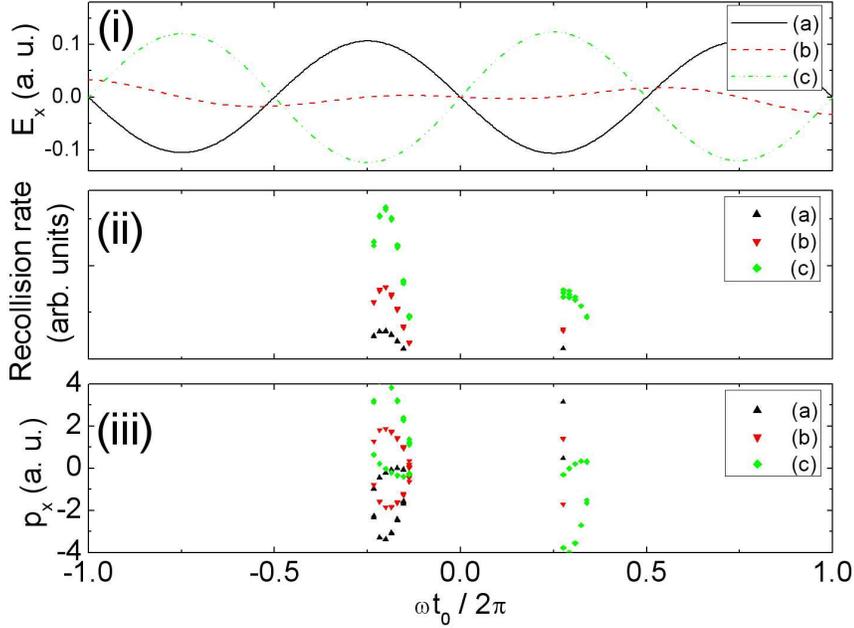}
\end{center}
\caption{Laser electric field in the x direction (panel (i)),
recollision rate (panel (ii)) and the final electron momentum along
$p_{1x}=p_{2x}=p_{x}$ (panel (iii)), as functions of the phase
$\omega t_{0}$, for a pulse described by Eqs. (\ref{eq3}) and
(\ref{eq4}), with CE phase of $0.5\pi$ and three different phase
delays of $\omega T_{d}=8\pi$, $7\pi$ and $6\pi$[tags (a), (b) and
(c), respectively].
The remaining parameters are the same as in Fig.~\ref{fig1}.}%
\label{frandPS_odddelay}%
\end{figure}

Still, one remaining question is how to understand the dramatic
change in the electron momentum distributions between the delay
phase of $8\pi$ and $6\pi$, as shown in Fig. \ref{fig1}. For that
purpose, we further conduct the calculations for the CE phase
$\phi=0.5\pi$ and several delay phases between $8\pi$ and $6\pi$,
i.e., $\omega T_{d}=7.5\pi,7\pi$ and $6.5\pi$. The results are
displayed in Fig. \ref{pd_odddelay}. In addition to the momentum
components $(p_{1x},p_{2x})$ in the x direction, we also consider
the momentum components $(p_{1y},p_{2y})$. The distributions as
functions of the former or latter momentum components are displayed
in the upper and lower panels of the figure, respectively.
Therefore, for the sake of clarity, we are no longer employing the
notation $p_{n\parallel}$.

It is found that, actually, the electron momentum distributions
evolve gradually from the third quadrant to the first quadrant, with
the decrease of the delay phase between $8\pi$ and $6\pi$. And
interestingly, at a delay phase of $7\pi$, the electron distribution
is almost centered at the origin (Fig.~\ref{pd_odddelay}(c)). This
is quite different from the case with a linearly polarized few-cycle
pulse, for which the center of the electron momentum distribution is
always at nonvanishing momenta and the distributions just ``jump"
from one quadrant to another. This is due to the fact that, in the
linearly-polarized, few-cycle case, the ``jump" is caused by a shift
in the dominant set of orbits, while, in the present situation, the
dominant set of trajectories remains the same. The changes occur in
the momentum components in the x direction.

This modification in the most probable momenta is due to the fact
that the polarization gate changes direction within this phase-delay
interval. In fact, as the x momentum components $p_{nx}, (n=1,2)$
decrease, there is a corresponding increase in the momentum
components $p_{ny}, (n=1,2)$. This is explicitly shown in the lower
panels of Fig.~\ref{pd_odddelay}, which exhibit exactly the opposite
behavior as in the upper panel, i.e., an increase in the peak
momenta of the distributions as $\omega T_d$ is shifted from $8\pi$
to $7\pi$, followed by a decrease when this delay is further shifted
to $6\pi$. In fact, while for Figs.~\ref{pd_odddelay}(f) and
\ref{pd_odddelay}(j) the distributions exhibit a peak near
$p_{1y}=p_{2y}=0$, in the central panel [Fig.~\ref{pd_odddelay}(h)]
the absolute value of the peak momentum is near $2\sqrt{U_p}$.

In order to understand this more clearly, we plot in Fig.
\ref{frandPS_odddelay} the laser electric field in the x direction,
the recollision rate and the final electron momentum along
$p_{1x}=p_{2x}=p_{x}$ as functions of the phase $\omega t_{0}$, for
three different delay phases of $\omega T_{d}=8\pi$, $7\pi$ and
$6\pi$. From the upper panel in Fig. \ref{frandPS_odddelay}, one
finds that, for $\omega T_{d}=7\pi$, the field amplitude in the x
direction is very small around the center of the gate, because the
two combining field components in x direction are totally out of
phase. While for $\omega T_{d}=8\pi$ and $6\pi$, the fields have
maximal amplitudes but opposite directions due to their relative
phase shift of $\pi$.

As discussed in section \ref{model}, the center of the hypersphere
which delimits the momentum distributions in momentum space is
determined by the vector potential-component $A_{x}(t)$ at the
instant when the electron recollides. For $\omega T_{d}=7\pi$, the
amount of momentum the electron acquired from the combined field is
thus very small in the x direction. Hence, the classically allowed
region is approximately centered at momentum $p_{x}=0$, as shown in
Fig.~\ref{frandPS_odddelay}(iii). For $\omega T_{d}=8\pi$ and
$6\pi$, the relative shift of $2\pi$ in the delay phase results in
the inversion of the x-component of the combined field at the very
center of the polarization gate. This gives rise to electron
emission in the opposite direction, i.e., the electron momentum
distribution shifts from the third to the first quadrant.

\section{Conclusions}
\label{conclusions}

In this work, we have studied laser-induced nonsequential double ionization
(NSDI) with a polarization-gated driving pulse. The specific pulse employed
here consisted of two few-cycle pulses with opposite circular polarizations
and a time delay $T_{d}$ (for details we refer to \cite{Chang2004}). The
ellipticity of the combined field is time-dependent and vanishes for $t_{0}%
=0$. The larger the time delay between both pulses, the steeper the
time-dependence in the polarization. We performed such studies
within a classical framework, extending the model in Ref.
\cite{LF2004,FLSL2004,FSLB2004} to elliptically polarized fields.

We found that the electron-momentum distributions, as functions of
the electron components parallel to the laser-field polarization at
the center of the gate, are very much dependent on the delay between
the two few-cycle pulses $\overrightarrow{E}_{r}$ and
$\overrightarrow{E}_{l}$ composing the polarization-gated pulse. For
long delays that are chosen as multiple times of the pulse duration,
the distributions are asymmetric and dependent on the CE phase of
such pulses. This is similar to the behavior observed for a single
linearly polarized, few-cycle driving pulse \cite{LF2004,FLSL2004}.
As the time delay $T_d$ decreases, this asymmetry fades. We could
explain this behavior in terms of trajectories. Below we will
summarize the main aspects of this explanation, and also draw a
parallel between the present situation and the previously studied
case of linearly polarized, few-cycle pulses \cite{LF2004,FLSL2004}.

For linearly polarized few cycle pulses, the momentum region in
which the electron momentum distributions will be concentrated
depends on the quasi-static tunneling rate for the first electron
and on the momentum region for which NSDI is classically allowed. A
large rate and a large momentum region imply that the contributions
from a specific set of trajectories to the distributions will be
prominent. For very short pulses, there is in general a single set
of trajectories that best fulfills such conditions. This set will
change with the CE phase. As the pulse length increases, more and
more sets of trajectories will lead to prominent contributions, and
the asymmetry will fade.

For polarization gated pulses, the most relevant set of orbits will be that
closest to the polarization gate, i.e., to the quasi linearly polarized region
of the pulse. Indeed, the first electron will tunnel more efficiently than for
the remaining set of orbits. This is due to the fact that the electron
velocity components perpendicular to the main polarization axis of the laser
field is smallest for the dominant set of orbits.

Apart from that, the first electron will return more easily to the parent ion
if it was released near the polarization gate. This will happen for two main
reasons. First, the small lateral velocity components will not lead to a
significant initial motion in the transverse direction. Second, the electron
will propagate in the continuum in a time interval close to the gate. Hence,
it will be accelerated in the transverse direction to a lesser extent than for
the remaining sets of orbits. This will guarantee an efficient return to the ion.

For an efficient polarization gate, the changes in the driving-field
ellipticity within the pulse length are very steep, so that only the orbits
close to the gate will contribute significantly to the distributions. In
general, this will lead to highly asymmetric electron momentum distributions,
concentrated in the momentum region determined by the dominant orbits. This is
a similar behavior to that encountered for linearly polarized driving fields,
in the sense that, in general, only one set of trajectories dominates the yield.

If the gate, however, is less efficient, the contributions from the
other sets of orbits will be increasingly important. Ultimately,
this will lead to nearly symmetric electron-momentum distributions.
For the specific driving pulse in this work, this effect has been
observed by decreasing the delay $T_{d}$ between the pulses.

In general, the efficient use of polarization gates could
tremendously increase the degree of attainable control of strong
field ionization dynamics, compared to the commonly used single
linearly polarized few-cycle pulse. For the latter, only the scalar
properties of laser pulses is used, while a polarization gate can,
in principle, as demonstrated in this work, make full use of the
flexibility of the vectorial laser field, e.g., the polarization
state, as an additional knob, to steer the NSDI dynamics within
sub-cycle time resolution.

\ack X. Liu acknowledges the financial support from the NNSF of
China (No. 10674153) and C.F.M.F from the UK EPSRC (Advanced
Fellowship, Grant no. EP/D07309X/1).

\end{document}